\documentclass[cernpreprint, atlasdraft=false, USenglish, texmf, orcidlogo]{atlasdoc}
\usepackage[subcaption]{atlaspackage}
\usepackage{atlasbiblatex}

\usepackage{atlasphysics}
\graphicspath{{logos/}{figures/}}

\usepackage{ANA-STDM-2021-01-PAPER-defs}

\AtlasTitle{Search for the low-mass isotropic signature of QCD instantons in $pp$ collisions at $\sqrt{s} = 13$~TeV using the ATLAS detector}

\AtlasAbstract{%
This Letter presents the first search for QCD instanton-induced events at the LHC\@.
Instantons are semi-classical solutions of the Yang--Mills equations of QCD and correspond to non-perturbative tunneling between topologically distinct vacuum sectors.
The search is performed in $1.6~\text{nb}^{-1}$ of very low instantaneous luminosity proton--proton collision data collected by the ATLAS detector at 13~TeV.
Angular variables based on reconstructed charged-particle tracks are used to discriminate the expected rest-frame isotropy of low-mass instanton events from the minimum-bias background.
The minimum-bias event background is predicted using a data-driven technique and reweighted Monte Carlo models.
A profile-likelihood fit is performed in a two-dimensional classifier variable to extract a possible signal contribution, yielding an excess of low-mass isotropic events with an observed significance of $3.5\sigma$.
}

\AtlasRefCode{STDM-2021-01}

\PreprintIdNumber{CERN-EP-2026-284}

\AtlasJournal{\PRL}
\hypersetup{pdftitle={ATLAS document},pdfauthor={The ATLAS Collaboration}}

\begin{document}

\maketitle

The Standard Model (SM) of particle physics is one of the most precisely tested theories in science.
However, several of its fundamental predictions related to the topological structure of non-Abelian gauge fields remain largely unexplored.
In particular, instantons are predicted in the SM as semi-classical solutions of the Euclidean-space Yang--Mills equations~\cite{1975-BPST-Pseudoparticle_solutions,1976-tHooft-Symmetry_breaking}, and correspond to quantum mechanical tunneling between topologically distinct vacuum states~\cite{1976-Callan_Dashen_Gross-gauge_theory_vacuum,1976-Jackiw_Rebbi-Vacuum_Periodicity}.
This multi-vacuum structure in Quantum Chromodynamics (QCD) gives rise to the $\theta$-vacuum and the strong CP problem~\cite{PhysRevLett.38.1440,Weinberg:1977ma,Wilczek:1977pj}.
Instantons are hypothesized to induce a rich non-perturbative phenomenology~\cite{1976-tHooft-Computation,1986-tHooft-U1,Shuryak:1981ff,Shuryak:1981fza,Diakonov:1983hh,Schafer:1995pz,Shifman:1978bx,Shuryak:1992ke,Diakonov:1995qy}, which remains empirically unverified despite decades of theoretical and experimental study.

Instanton field configurations contribute to path integral dynamics through the 't~Hooft vertex~\cite{1976-tHooft-Computation,1986-tHooft-U1}.
This effective interaction must involve a quark--antiquark pair of each active flavor due to an index theorem associated with the vertex.
The instanton is an extended field configuration in Euclidean space and is characterized by its size, $\rho$.
Its momentum scale is set by $1/\rho$, so it can only resolve a quark of mass $m_f$ if $\rho < 1 / m_f$ according to the wavelength argument of Refs.~\cite{2020-Khoze-Large_effects,Mangano:2021udl}, while the instanton mass is defined as the center-of-mass energy of the transition.
Since the instanton has no preferred orientation in Euclidean space, it is reasonable to assume a distinctive, approximately isotropic distribution of final-state particles.

The parton-level cross-section of the QCD instanton process at the LHC is obtained in Ref.~\cite{2020-Khoze-Large_effects} based on a semi-classical approximation of the QCD path integral for the gluon-induced transition $gg \rightarrow u\bar{u}d\bar{d}s\bar{s}c\bar{c} + \text{gluons}$, building on the methodology of Refs.~\cite{Ringwald:1989ee,Espinosa:1989qn,Khoze:1990bm,Khoze:1991mx} with leading quantum corrections from Refs.~\cite{Mueller:1991fa,Mueller:1991fp}.
The calculation predicts a cross-section that falls steeply with instanton mass, and a mean final-state gluon multiplicity $\langle N_g\rangle$ that increases from 5.5 to 6.8 for instanton masses in the range 20 to 50~\GeV, while $\langle 1/\rho\rangle$ varies from 1.6 to 3.1~\GeV.
Below around 20~\GeV in mass, the semi-classical approximation and the corresponding cross-section predictions become increasingly unreliable~\cite{2020-Khoze-Large_effects,Guin:2026bsq}.
The uncertainty in the cross-section spans orders-of-magnitude due to the exponential dependence of the amplitude on the strong coupling constant $\alphas$ and sensitivity to the choice of hard scale~\cite{Mangano:2021udl}.

The experimental challenge is to disentangle the rest-frame isotropy of QCD instanton events from soft QCD processes including multiple-parton interactions (MPI).
Phenomenological studies have proposed using event shape observables~\cite{2020-Khoze-Large_effects,2021-Schott}, and large rapidity gaps or forward proton tags in the diffractive production channel~\cite{2021-Khoze-Hunting_for_QCD_Instantons,Tasevsky2023}.
Previous searches were performed in electron--proton collisions by the H1~\cite{H1:2002lra,H1:2016jnv} and ZEUS~\cite{ZEUS:2003qkm} experiments at HERA based on similar predictions~\cite{Ringwald:1994kr,Moch:1996bs,Ringwald:1998ek}.

This Letter presents the first search for QCD instanton-induced events at the Large Hadron Collider (LHC).
The search is performed in \SI{1.6}{\per\nano\barn} of minimum-bias data collected at a proton--proton ($pp$) center-of-mass energy of $\sqrt{s} = \SI{13}{\TeV}$~\cite{STDM-2015-02,STDM-2016-07,STDM-2018-60}.
The analysis relies on the expected experimental signature of a low-mass isotropic decay, and exploits the high granularity of the ATLAS inner detector (ID) to reconstruct angular correlation variables from charged-particle tracks.
A deep neural network (DNN) is used to discriminate the signal from the minimum-bias background, and to facilitate a data-driven estimation of the background.
Reweighted Monte Carlo (MC) models are used to quantify systematic uncertainties on the result.

The ATLAS detector~\cite{PERF-2007-01,SOFT-2010-01} is a general-purpose particle detector at the LHC\@.
The ID~\cite{IDET-2010-01,PERF-2015-07,IDTR-2019-05,IDET-2020-01} provides charged-particle tracking over the pseudorapidity range $|\eta| < 2.5$~\cite{atlas-footnote}, combining a silicon pixel detector~\cite{IDET-2017-10,PIX-2018-001,PIXE-2023-02}, a silicon microstrip tracker (SCT)~\cite{SCTD-2019-01}, and a transition radiation tracker (TRT)~\cite{IDET-2015-01}, all immersed in a 2~T solenoidal magnetic field.
Minimum-bias events are selected~\cite{TRIG-2019-04} by requiring at least one hit in the minimum-bias trigger scintillators~\cite{ATLAS-CONF-2010-068,Sidoti_2014}.
The data were collected in 2015 under very low instantaneous luminosity LHC conditions, where the mean number of $pp$ interactions per bunch crossing (pileup) was $\langle\mu\rangle = 0.028$~\cite{DAPR-2018-01} to minimize the background from pileup interactions.

Charged-particle tracks are reconstructed from hits in the ID and are required to satisfy the Tight Primary selection working point~\cite{ATL-PHYS-PUB-2015-051} to ensure high purity of prompt charged particles while suppressing the contribution of combinatorial fake tracks.
Primary vertices (PVs) are fitted from the reconstructed tracks and require at least two tracks with transverse momenta $\pT > \SI{100}{\MeV}$~\cite{ATL-PHYS-PUB-2015-026}, and are ordered by the sum of squared transverse momenta of the associated tracks.
The event selection follows \Refn{\cite{STDM-2015-02}} in requiring one reconstructed PV, and rejecting events that have a second PV with five or more associated tracks.
Tracks entering the analysis must have $\pT > \SI{500}{\MeV}$, $|\eta| < 2.5$, and be associated to the PV by requiring a transverse impact parameter relative to the beamline $|d_0^{\text{BL}}| < \SI{1.5}{\mm}$ and a longitudinal impact parameter $|z_0^\text{PV} \sin{\theta}| < \SI{1.5}{\mm}$ relative to the PV.

Instanton signal events are simulated using the \Sherpa{3.0.1} event generator~\cite{Sh3} as described in \Refn{\cite{2020-Khoze-Large_effects}}.
The input parton-level cross-section values from \Refn{\cite{2020-Khoze-Large_effects}} are interpolated to obtain a smooth instanton mass spectrum.
The final state contains the $u,\bar{u},d,\bar{d},s,\bar{s},c$ and $\bar{c}$ quarks, plus a number of additional gluons taken from a Poisson distribution of mean $\langle N_g\rangle$.
Events are hadronized using the cluster model~\cite{Winter:2003tt}.
The renormalization and factorization scales are set to the inverse instanton size, $\langle 1/\rho\rangle$.
Only instantons with masses above \SI{20}{\GeV} are generated~\cite{2020-Khoze-Large_effects}.
The signal event simulation uses the \CT[18nnlo] parton distribution function (PDF)~\cite{Hou:2019efy}.
The occurrence of multiple-instanton interactions within a single \pp collision is estimated to be negligible.
The total hadronic cross-section after these choices is about $\SI{0.23}{\milli\barn}$.

The \EPOS LHC event generator~\cite{Werner:2005jf,Pierog:2013ria} is used to simulate minimum-bias background events.
As an alternative, \PYTHIA[8.230]~\cite{Sjostrand:2014zea,ATL-PHYS-PUB-2019-017} is used, with MPI simulated based on the Sj\"ostrand-van Zijl model~\cite{1987-Sjostrand-MPI,Sjostrand:2004pf} and the underlying event and parton shower parameters set to the A14 tune~\cite{ATL-PHYS-PUB-2014-021}.
\PYTHIA is used with the \NNPDF[2.3lo] PDF set~\cite{Ball:2012cx} and Lund string fragmentation~\cite{Sjostrand:1984ic}.
A third background sample is generated from the same version and configuration of \Sherpa as was used for the signal, except with Lund string fragmentation.
The response of the ATLAS detector is simulated using the \GEANT toolkit~\cite{Agostinelli:2002hh} within the ATLAS simulation infrastructure~\cite{SOFT-2010-01}.

\begin{figure}[t]
\centering
\begin{subfigure}{0.32\textwidth}
\centering
\includegraphics[width=\textwidth]{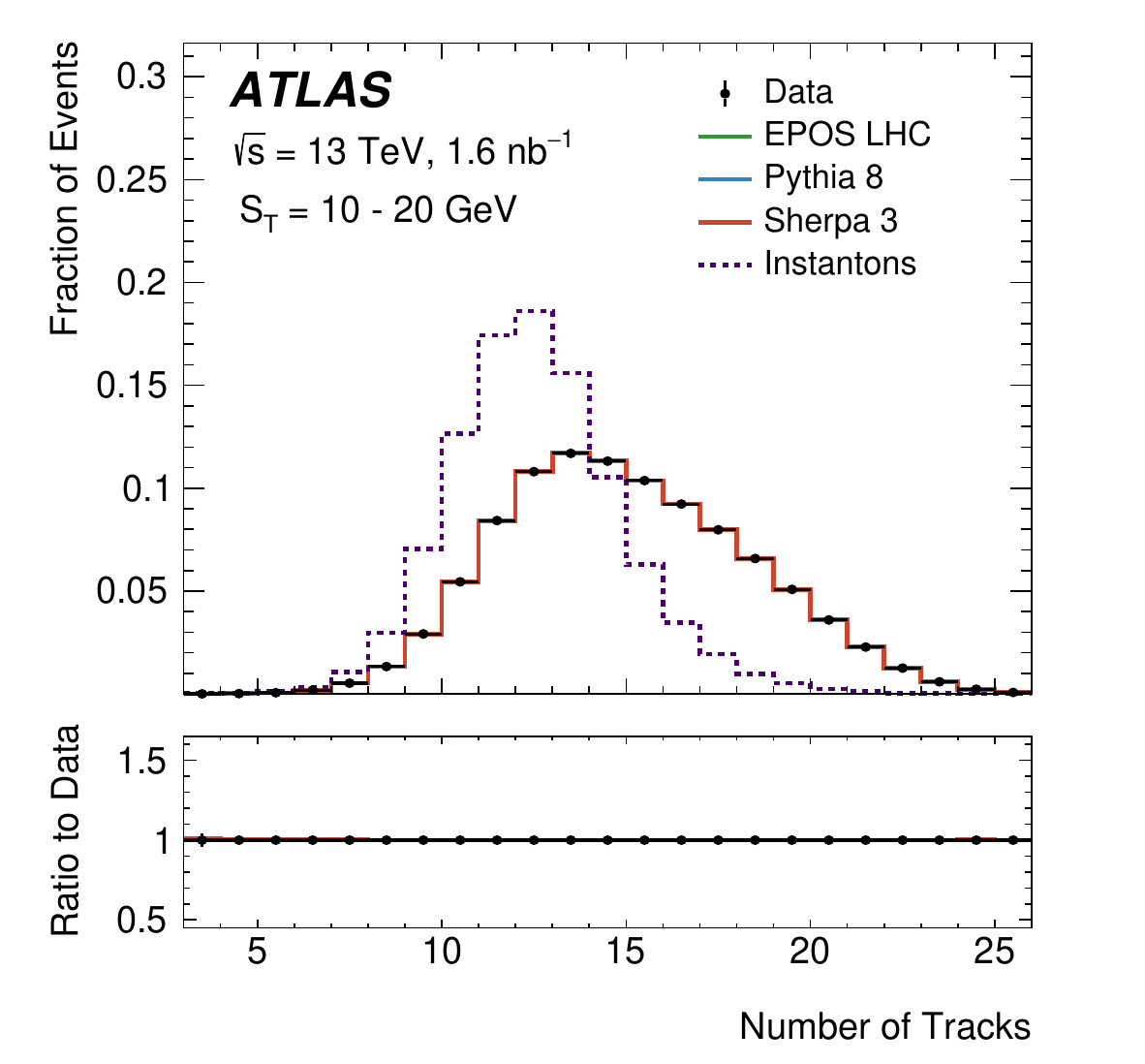}
\caption{}
\end{subfigure}
\begin{subfigure}{0.32\textwidth}
\centering
\includegraphics[width=\textwidth]{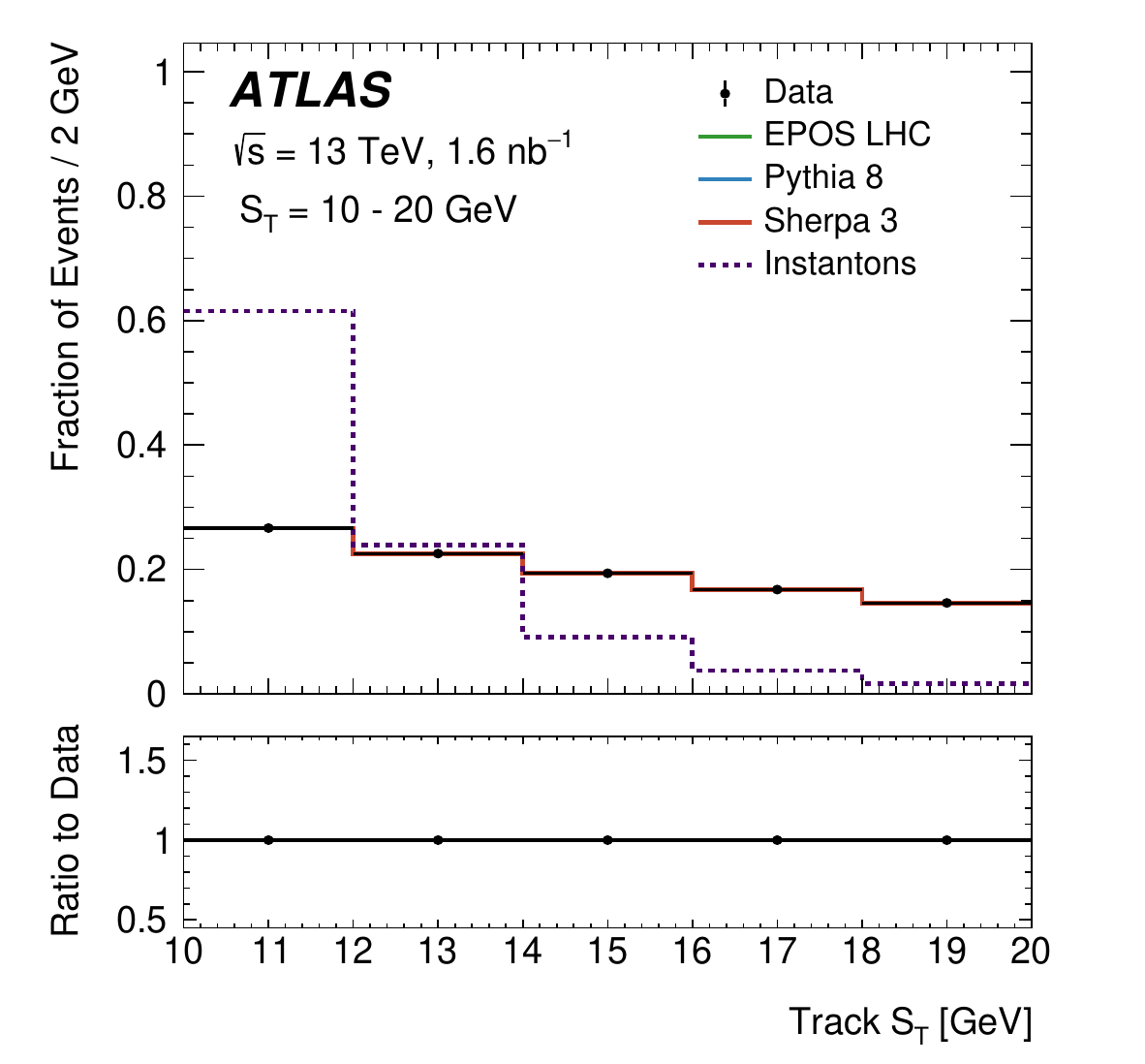}
\caption{}
\end{subfigure}\\
\begin{subfigure}{0.32\textwidth}
\centering
\includegraphics[width=\textwidth]{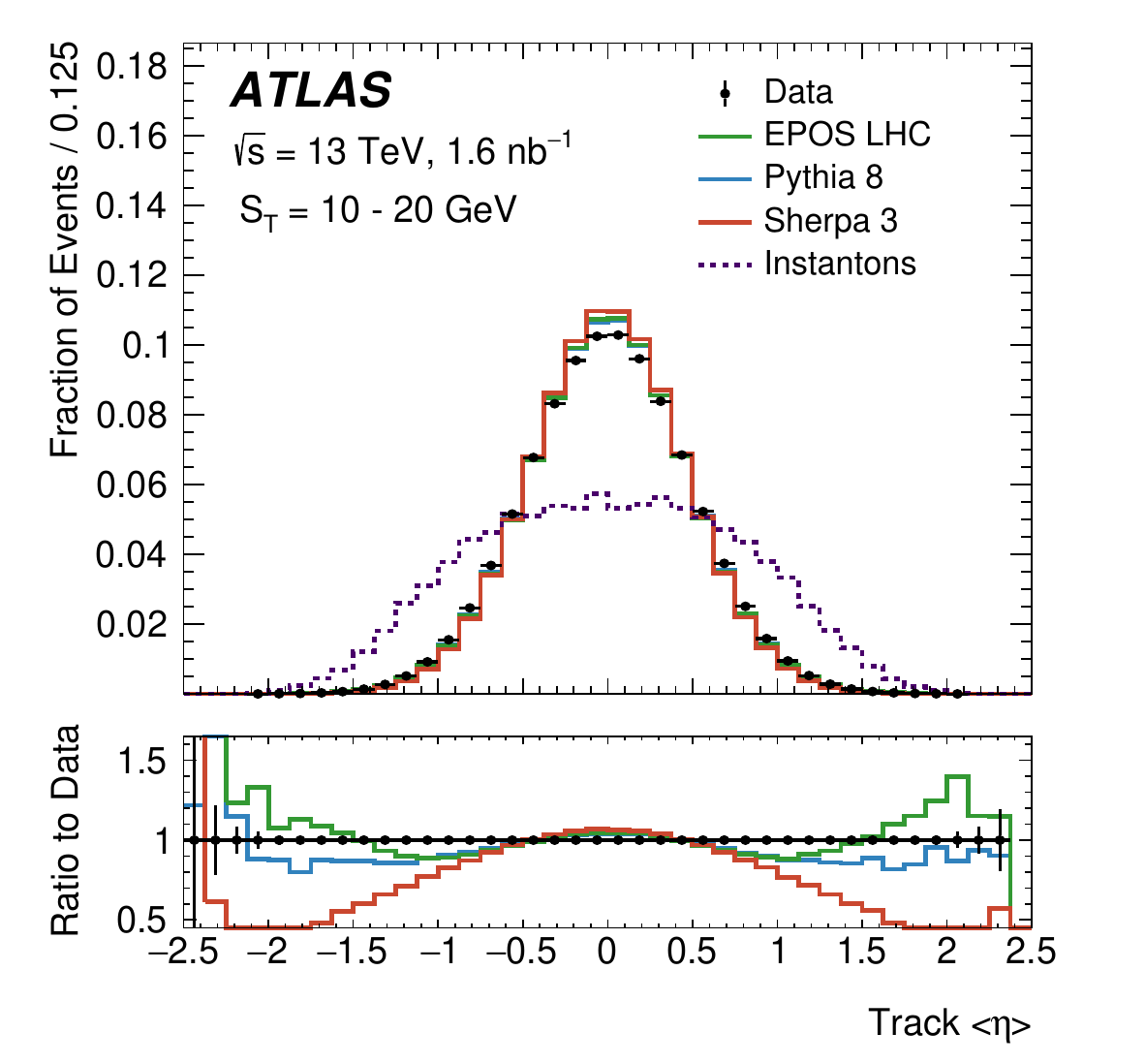}
\caption{}
\end{subfigure}
\begin{subfigure}{0.32\textwidth}
\centering
\includegraphics[width=\textwidth]{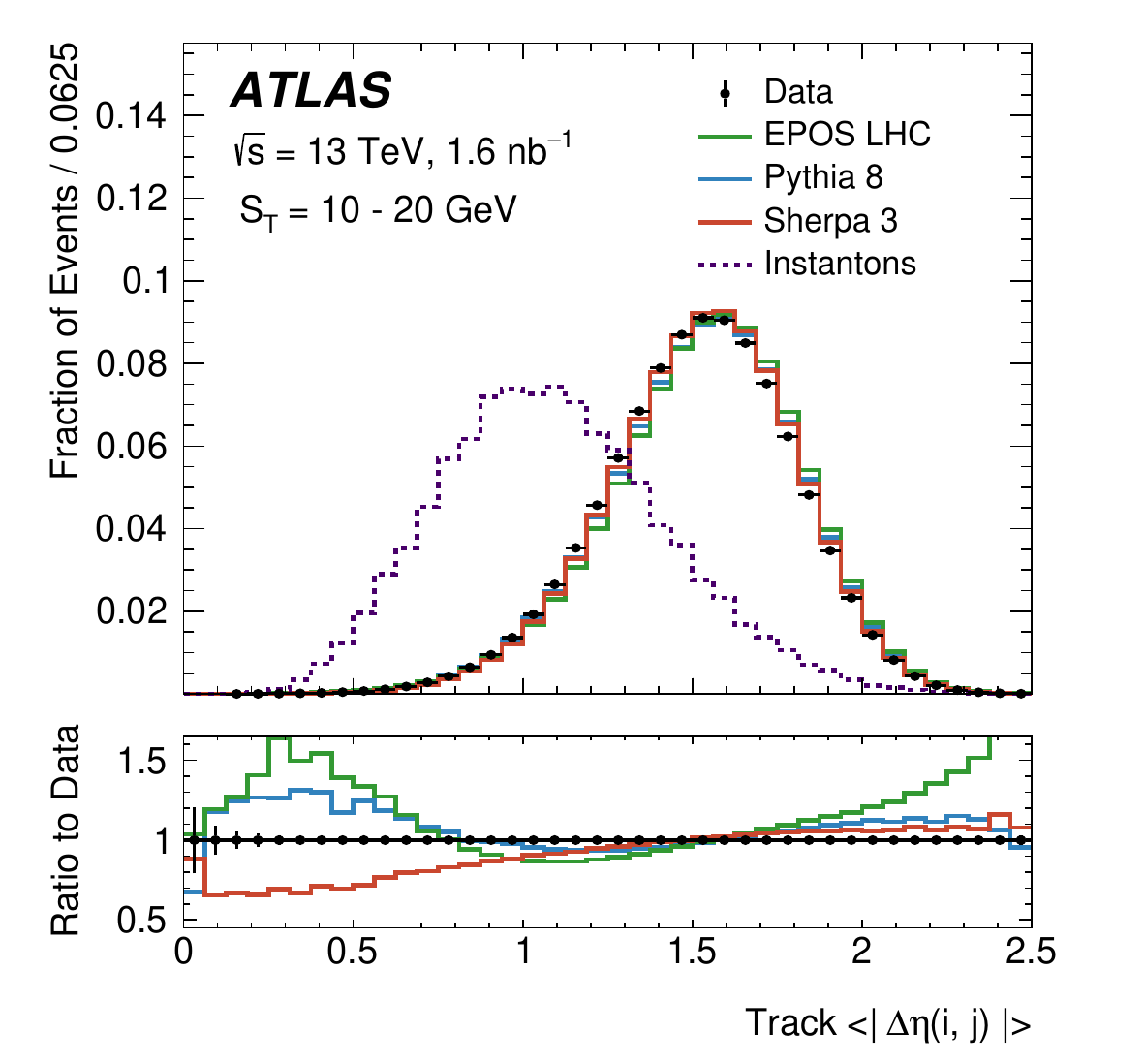}
\caption{}
\end{subfigure}
\begin{subfigure}{0.32\textwidth}
\centering
\includegraphics[width=\textwidth]{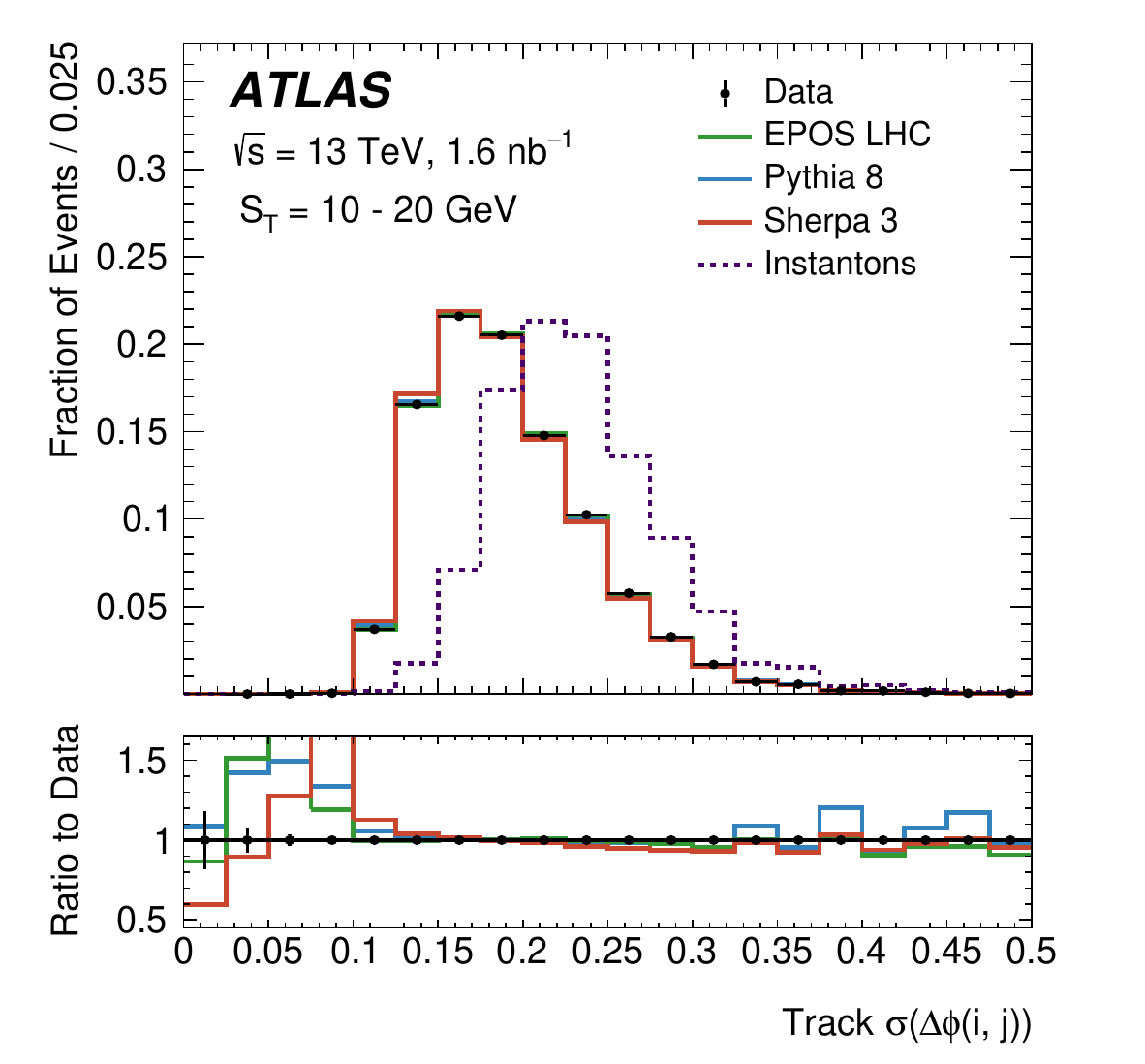}
\caption{}
\end{subfigure}
\caption{Unit-normalized distributions of the DNN inputs $N_\text{trk}$, $\ST$, $\langle\eta\rangle$, $\langle|\Delta\eta|\rangle$, and $\sigma(\Delta\phi)$ after selecting events with $10 < \ST < 20~\si{\GeV}$.  The three background MC samples \PYTHIA[8], \EPOS LHC, and \Sherpa[3] are reweighted to the data inclusively in $N_\text{trk}$ and $\ST$ as described in the text. The uncertainties on the data points are statistical only.}
\label{fig:inputs}
\end{figure}

For signal discrimination, five event-level observables are considered: the track multiplicity, $N_\text{trk}$; the scalar sum of track transverse momenta, $\ST$; the mean track pseudorapidity, $\langle\eta\rangle$; the mean pseudorapidity separation between every pair of tracks in the event, $\langle|\Delta\eta|\rangle$; and the standard deviation of the distribution of the azimuthal separation of all pairs of tracks in the event, $\sigma(\Delta\phi)$.
Event shape observables such as thrust and sphericity are not utilized because they are poorly modeled~\cite{STDM-2011-32,ALICE:2023bga,CMS-SMP-23-008}.

To improve the modeling of the data, the background MC models are reweighted simultaneously in two dimensions to the inclusive $N_\text{trk}$ and $\ST$ distributions of the data.
This corrects the shapes of the $N_\text{trk}$ and $\ST$ distributions while also fixing the total cross-sections of the background models to the data, and is applied to each of the MC models \EPOS, \PYTHIA{8}, and \SHERPA{3} henceforth.
A signal injection test is performed to verify that this inclusive reweighting does not affect sensitivity to the signal at the relevant level of contamination.

To mitigate the impact of very-low-mass instantons, the selection $10 < \ST < 20$~\GeV is imposed, which has an acceptance times efficiency of 7.4\% (27\%) for signal (background) events.
Furthermore, this reduces sensitivity to the predicted mass spectrum, effectively rendering signal theory shape uncertainties negligible.
The distributions of the signal and background MC samples in the five selected observables within the final analysis phase space are shown in \Fig{\ref{fig:inputs}}.
Contamination from lower mass instanton events is assessed using a lower mass sample, and is found to cause only a normalization effect.

The $N_\text{trk}$ and $\ST$ variables are sensitive to the hard scale of the event, while the $\sigma(\Delta\phi)$ variable quantifies the uniformity of the distribution of tracks in the azimuthal plane.
\Fig{\ref{fig:inputs}} shows that the signal process produces fewer tracks and a softer $\ST$ spectrum.
The $\langle|\Delta\eta|\rangle$ variable represents the width of the track pseudorapidity distribution event-by-event, and is smaller for the signal because of its isotropy and relatively low MPI activity.
In contrast, the independent parton--parton scatters of the MPI background process within the background models tend to populate the ID more uniformly, resulting in higher values of $\langle|\Delta\eta|\rangle$.
The $\langle\eta\rangle$ variable captures the longitudinal Lorentz boost of the pseudorapidity distribution and is broader for signal events.

The phenomenological nature of the background models and their dependence on tunable parameters motivates a data-driven approach to background estimation.
To this end, the ABCDisCo~\cite{2021-ABCDisCo} technique, used previously in Refs.~\cite{EXOT-2022-04,CMS-MLG-23-003}, is employed to construct two discriminating variables that are statistically independent of one another as a basis for ABCD background estimation~\cite{ATL-PHYS-PUB-2026-014}.
The ABCDisCo DNN is trained with the five kinematic variables $N_\text{trk}$, $\ST$, $\sigma(\Delta\phi)$, $\langle\eta\rangle$, $\langle|\Delta\eta|\rangle$ as inputs, and has two output nodes $X$ and $Y$.
The training is governed by a modified loss function $\mathcal{L} = \mathcal{L}_\text{bce}[X] + \mathcal{L}_\text{bce}[Y] + \alpha\cdot\text{DisCo}[X, Y]$, where the first two terms correspond to binary cross-entropy (bce) classification terms that apply to the output nodes $X$ and $Y$, respectively, while the last term represents the distance correlation between them.
Distance correlation is a non-linear measure of correlation that takes the value zero (one) for statistically independent (dependent) random variables~\cite{2007-Szekely-dcorr,2020-DisCo_Fever}.
The $\alpha$ coefficient controls the strength of the decorrelation term and is optimized simultaneously with the DNN hyperparameters.
The DNN is trained with the reweighted \EPOS prediction as background, and the \Sherpa instanton MC as signal.
The DisCo term is only applied to the background.
It is verified that the DNN discriminates the properties of the instanton final state itself as opposed to the details of the \Sherpa underlying event or hadronization models.
Appendix~\ref{app:dnn} further details the DNN setup.

The two ABCDisCo classifier variables $X$ and $Y$ are used to define a two-dimensional plane in which the ABCD background estimation is performed~\cite{ATL-PHYS-PUB-2026-014}.
The ABCD method relies on the two classifier variables being statistically independent for the background process.
However, residual correlations arise due to imperfect decorrelation performance at high classifier values and the DisCo penalty term allowing local correlations.
These residual correlations are removed by reweighting the ABCD plane bin-by-bin by a non-closure factor $w_{ij} = N_{ij}^\text{MC~ABCD} / N_{ij}^\text{MC}$, where $i$ and $j$ are bin indices in classifiers $X$ and $Y$.
Here, $N_{ij}^\text{MC}$ is the event yield predicted by the MC, while $N_{ij}^\text{MC~ABCD}$ is the event yield predicted by the ABCD method applied to the same MC\@.
The nominal weights $w_{ij}$ are derived from the mean of the \EPOS, \PYTHIA{8}, and \Sherpa{3} background MC predictions~\cite{closure-footnote}, removing the non-closure in the mean MC by construction, but not necessarily in the individual MC predictions, which have different correlation structures in the ABCD plane.
An alternative reweighting based on \EPOS gives compatible results.

A likelihood fit is performed following the framework of Refs.~\cite{2011-Cowan-Asymptotic_formulae,TRExFitter} to constrain the parameters of the background model from the data and to extract any signal.
Systematic uncertainties are incorporated as nuisance parameters (NPs) in the likelihood function with associated Gaussian constraint terms.
The parameter of interest is the normalization of the signal, called the signal strength, $\mu$, which is unconstrained in the likelihood function.
Control, validation, and signal regions (SR) are defined based on the signal-to-background ratio, $S/B$, as given by the signal MC prediction and the reweighted \EPOS MC\@.
The regions are optimized for the correct extraction of the signal based on signal-injection tests before unblinding the SR data and to be robust against signal contamination considering the large signal cross-section uncertainty.
The closure-weighted (unweighted) SR signal acceptance within the selected \ST interval is around 6.1\% (15\%).
Appendix~\ref{app:likelihood} details the likelihood fit and systematic uncertainties.

The likelihood fit yields an excess of events with an observed (expected) significance of $3.5\sigma$ ($1.5\sigma$) relative to the minimum bias background-only hypothesis.
The total post-fit event yield in the SR is $(\numRF[3]{2.878690} \pm \numRF[2]{0.109397}) \times 10^{4}$ for the background and $(\numRF[2]{3.94195} \pm \numRF[2]{1.07858}) \times 10^{3}$ for the signal, compared with \num{32789} data events.
The fitted number of signal events corresponds to a visible signal cross-section of $2.4 \pm 0.7~\si{\micro\barn}$ in the SR, where the uncertainty of 2.1\% in the integrated luminosity is included in the total uncertainty.
The fitted signal cross-section corresponds to scaling the predicted signal cross-section by $\mu = 2.3 \pm 0.6$.
The quoted uncertainty is experimental only, and is dominated by the uncertainty on the background estimation (see Appendix~\ref{app:likelihood}).
Uncertainties due to the track reconstruction enter through the detector simulation and affect the signal MC sample, but these have negligible impact on the signal extraction.
The model-independent upper limit on the visible cross-section is $3.51~\si{\micro\barn}$ ($1.41 \pm 0.03~\si{\micro\barn}$ expected) at 95\% confidence level based on the $\mathrm{CL}_{\mathrm{s}}$ method~\cite{Read:2002hq}.

The analysis methodology was tested in two higher-$S_\text{T}$ intervals with $20 < \ST < 30$~\GeV and $30 < \ST < 40$~\GeV, where the relative signal contribution is at least 30 times smaller than in the nominal \ST interval and therefore negligible.
The likelihood fit returns a signal strength compatible with zero in both cases and gives a robust description of the data.

\begin{figure}[t]
\centering
\includegraphics[width=0.6\textwidth]{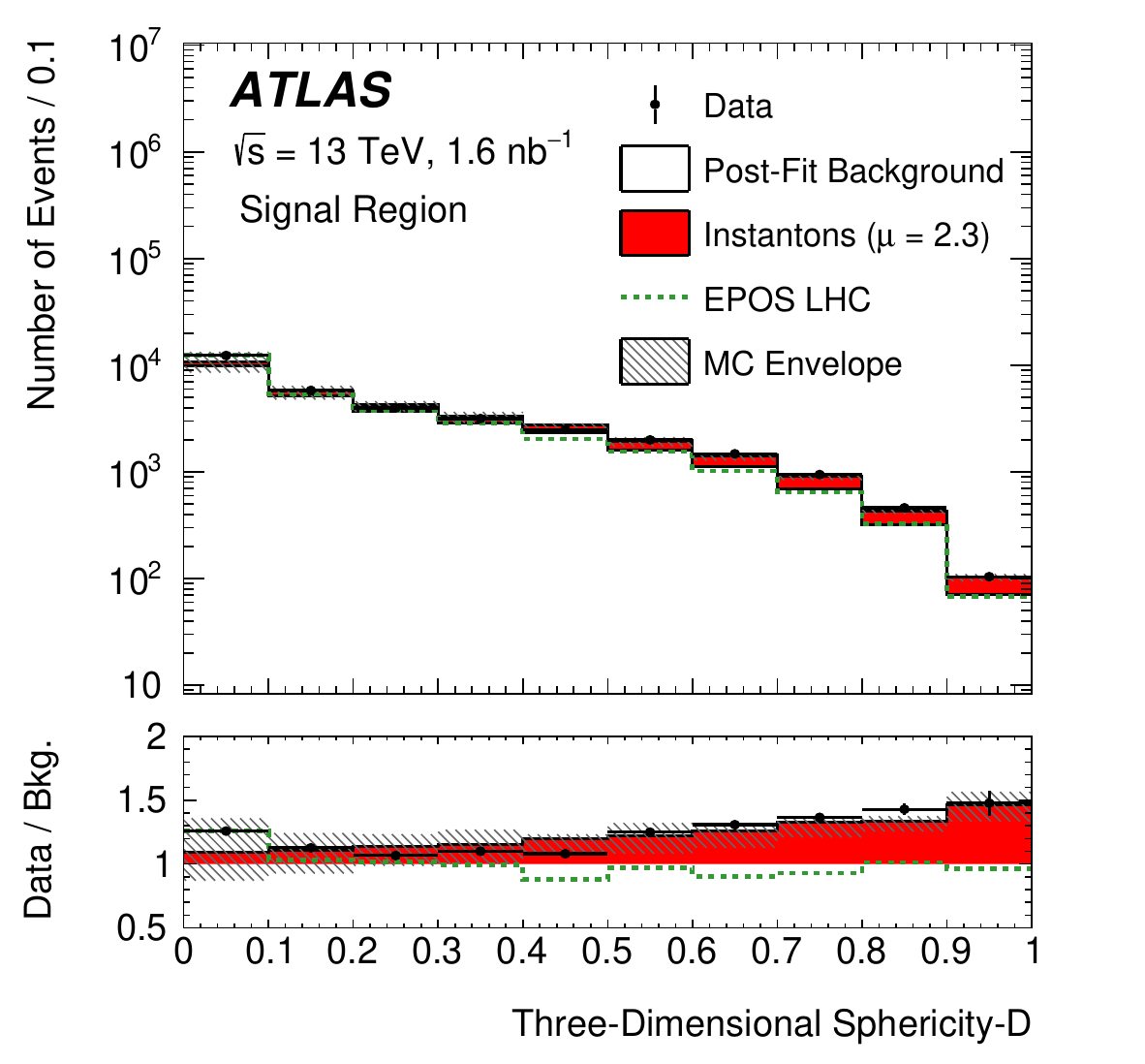}
\caption{Post-fit distribution of the three-dimensional sphericity-$D$ parameter in the signal region.  The post-fit background model is shown as the white histogram, with the instanton signal scaled by the best-fit signal strength $\mu = 2.3 \pm 0.6$ stacked on top.  The reweighted \EPOS LHC distribution is shown for comparison.  The envelope of the three reweighted background models \EPOS, \Sherpa{3}, and \PYTHIA{8} is combined in quadrature with the post-fit uncertainties on the fitted NPs to form the hatched band.  The data and the signal MC event yields are weighted by the ABCD non-closure correction factors.  The uncertainties on the data points are statistical only.}
\label{fig:postfit}
\end{figure}

The observed excess is characterized by low track multiplicity, broad $\langle\eta\rangle$, and relatively small $\langle|\Delta\eta|\rangle$.
\Fig{\ref{fig:postfit}} presents the post-fit SR distribution of the three-dimensional sphericity-$D$ parameter~\cite{sphericity-footnote}.
To obtain this, the fitted background model is projected onto sphericity-$D$ by assuming the correlations of the mean MC with the classifier.
At higher values of sphericity-$D$ corresponding to greater event isotropy, the data show a systematic excess compatible with the signal hypothesis.

\FloatBarrier

In summary, this Letter presents the first direct search for QCD instantons at the LHC\@.
The search is performed in $1.6~\text{nb}^{-1}$ of $\sqrt{s} = 13~\text{\TeV}$ $pp$ collision data collected by the ATLAS Experiment.
An excess of low-mass isotropic events is observed with a significance of $3.5\sigma$.
The excess is characterized as broader in mean pseudorapidity in comparison to the soft QCD background processes, is localized in a relatively narrow band in pseudorapidity within each event, and is isotropic in terms of the sphericity-$D$ event shape variable.
No background-only model can fully describe the data.
The interpretation of the excess is limited by the modeling of the background, highlighting the need for improved understanding of low-energy QCD processes.
Further characterization of the heavy-flavor signature and mass spectrum of the signal process is required to verify the instanton hypothesis.
This result is an important step in the experimental search for QCD instantons and the exploration of the topological structure of the Standard Model.

\section*{Acknowledgments}
%

%

% The next lines are included from the .//acknowledgements/Acknowledgements.tex input file
%
%

%
%

We thank CERN for the very successful operation of the LHC and its injectors, as well as the support staff at
CERN and at our institutions worldwide without whom ATLAS could not be operated efficiently.

The crucial computing support from all WLCG partners is acknowledged gratefully, in particular from CERN, the ATLAS Tier-1 facilities at TRIUMF/SFU (Canada), NDGF (Denmark, Norway, Sweden), CC-IN2P3 (France), KIT/GridKA (Germany), INFN-CNAF (Italy), NL-T1 (Netherlands), PIC (Spain), RAL (UK) and BNL (USA), the Tier-2 facilities worldwide and large non-WLCG resource providers. Major contributors of computing resources are listed in Ref.~\cite{ATL-SOFT-PUB-2026-001}.

We gratefully acknowledge the support of ANPCyT, Argentina; YerPhI, Armenia; ARC, Australia; BMWFW and FWF, Austria; ANAS, Azerbaijan; CNPq and FAPESP, Brazil; NSERC, NRC and CFI, Canada; CERN; ANID, Chile; CAS, MOST, NSFC and FRFCU, China; Minciencias, Colombia; MEYS CR, Czech Republic; DNRF and DNSRC, Denmark; IN2P3-CNRS and CEA-DRF/IRFU, France; SRNSFG, Georgia; BMFTR, HGF and MPG, Germany; GSRI, Greece; RGC and Hong Kong SAR, China; ICHEP and Academy of Sciences and Humanities, Israel; INFN, Italy; MEXT and JSPS, Japan; CNRST, Morocco; NWO, Netherlands; RCN, Norway; MNiSW, Poland; FCT, Portugal; MNE/IFA, Romania; MSTDI, Serbia; MSSR, Slovakia; ARIS and MVZI, Slovenia; DSI/NRF, South Africa; MICIU/AEI, Spain; SRC and Wallenberg Foundation, Sweden; SERI, SNSF and Cantons of Bern and Geneva, Switzerland; NSTC, Taipei; TENMAK, T\"urkiye; STFC/UKRI, United Kingdom; DOE and NSF, United States of America.

Individual groups and members have received support from BCKDF, CANARIE, CRC and DRAC, Canada; CERN-CZ and PRIMUS, Czech Republic; COST, ERC, ERDF, Horizon 2020 and Marie Sk{\l}odowska-Curie Actions, European Union; Investissements d'Avenir Labex, Investissements d'Avenir Idex and ANR, France; DFG and AvH Foundation, Germany; Herakleitos, Thales and Aristeia programmes co-financed by EU-ESF and the Greek NSRF, Greece; BSF-NSF and MINERVA, Israel; NCN and NAWA, Poland; La Caixa Banking Foundation, CERCA and AGAUR programs from Generalitat de Catalunya and PROMETEO and GenT Programmes Generalitat Valenciana, Spain; G\"{o}ran Gustafssons Stiftelse, Sweden; The Royal Society and Leverhulme Trust, United Kingdom; Eric and Wendy Schmidt Fund for Strategic Innovation, United States of America.

In addition, individual members wish to acknowledge support from Chile: Agencia Nacional de Investigaci\'on y Desarrollo (FONDECYT 1230812, FONDECYT 1230987, FONDECYT 1240721, FONDECYT 1240864, FONDECYT 3240661); China: Fundamental Research Funds for the Central Universities (010-63263105), Chinese Ministry of Science and Technology (MOST-2023YFA1605700), National Natural Science Foundation of China (NSFC-W2543003, NSFC-W2543005); Czech Republic: Czech Science Foundation (GACR - 24-11373S), Ministry of Education Youth and Sports (ERC-CZ-LL2327, FORTE CZ.02.01.01/00/22\_008/0004632); EU: H2020 European Research Council (ERC - 101002463); European Union: European Research Council (BARD No. 101116429, ERC - 101219398, ERC 101089007), European Regional Development Fund (HE COFUND GA No.101081355, ERDF), Marie Sklodowska-Curie Actions (GAP-101168829); France: Agence Nationale de la Recherche (ANR-22-EDIR-0002, ANR-24-CE31-0504-01); China: Research Grants Council (GRF); Italy: Istituto Nazionale di Fisica Nucleare (LHC-MIUR - 28003/2025), Ministero dell'Università e della Ricerca (NextGenEU I53D23000820006 M4C2.1.1, SOE2024\_0000023); Japan: Japan Society for the Promotion of Science (JSPS KAKENHI  JP25H0063, JSPS KAKENHI JP22H04944, JSPS KAKENHI JP24K23939, JSPS KAKENHI JP24KK0251, JSPS KAKENHI JP25H00650, JSPS KAKENHI JP25H01291, JSPS KAKENHI JP25K01011, JSPS KAKENHI JP25K01023, JSPS KAKENHI JP25KK0047); Poland: Polish National Science Centre (NCN OPUS 2023/51/B/ST2/02507, NCN OPUS nr 2022/47/B/ST2/03059, UMO-2019/34/E/ST2/00393, UMO-2021/42/E/ST2/00350, UMO-2022/47/O/ST2/00148, UMO-2023/49/B/ST2/04085, UMO-2023/51/B/ST2/00920, UMO-2024/53/N/ST2/00869); Spain: Agència de Gestió d'Ajuts Universitaris i de Recerca. (AGAUR - 2023 BP 00141), Ministry of Science and Innovation (RYC2019-028510-I, RYC2020-030254-I, RYC2021-031273-I, RYC2022-038164-I), Ministerio de Ciencia, Innovación y Universidades/Agencia Estatal de Investigaci\'on (PID2022-142604OB-C22); Sweden: Carl Trygger Foundation (Carl Trygger Foundation CTS 22:2312), Swedish Research Council (Swedish Research Council 2023-04654, VR 2021-03651, VR 2022-03845, VR 2022-04683, VR 2023-03403, VR 2024-05451, VR 2025-05940), Knut and Alice Wallenberg Foundation (KAW 2023.0366); United Kingdom: The Binks Trust, Royal Society (NIF-R1-231091); United States of America: U.S. Department of Energy (ECA DE-AC02-76SF00515), John Templeton Foundation (John Templeton Foundation 63206), Neubauer Family Foundation.

%
%

% End of text imported from the .//acknowledgements/Acknowledgements.tex input file

%
\clearpage
\appendix
\part*{End Matter}

\section{Details of the Deep Neural Network}\label{app:dnn}

The DNN is implemented using the Keras framework~\cite{chollet2015keras} with the TensorFlow backend~\cite{2015-tf}.
The optimal DNN architecture is found to consist of two layers of 128 hidden nodes each, with a drop-out rate of 0.25 between the layers, comprising \num{17666} trainable parameters, and $\alpha = 15$.
Large batch sizes of \num{10000} events are used during the training to enhance the statistical accuracy of the DisCo calculation.
The values of the loss components encountered at the point of minimum total loss in the validation dataset during training are $\mathcal{L}_\text{bce}[X] = 0.03504$, $\mathcal{L}_\text{bce}[Y] = 0.03174$, and $\text{DisCo}[X, Y] = 0.0001$.
This classification performance translates into an area under the ROC curve (AUC) of 0.82 and 0.85 for classifiers $X$ and $Y$, respectively.

\section{Details of the Background Model and Likelihood Fit}\label{app:likelihood}

\begin{figure}
\centering
\includegraphics[width=0.8\textwidth]{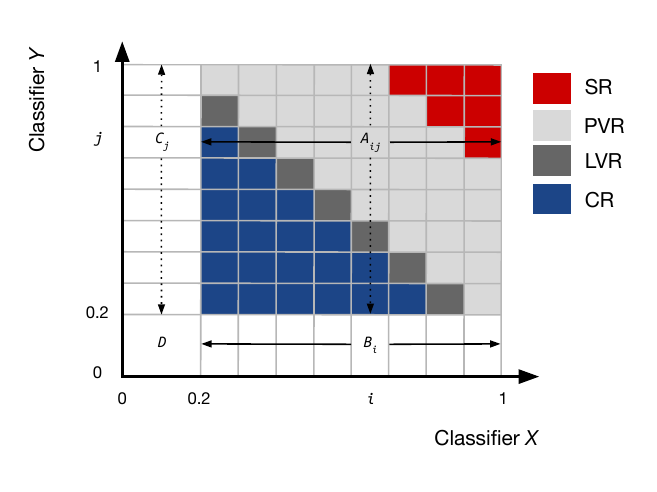}
\caption{Illustration of how the ABCD background estimation is performed in the ABCD plane.  For each bin with indices $(i, j)$, the background is estimated according to $A_{ij} = (B_i \times C_j)/D$, where $B_i$, $C_j$, and $D$ are the bin yields indicated.  The arrows around bin $(i, j)$ show how this method generates a binned ABCD background estimation across the A region.  Within the A region, the control, likelihood validation, physics validation, and signal regions used in the likelihood fit are shown in the solid colors.}
\label{fig:region_definitions}
\end{figure}

The ABCD method for the background estimate in the two-dimensional plane spanned by Classifiers $X$ and $Y$ is illustrated in \Fig{\ref{fig:region_definitions}} together with the definitions of the regions used in the likelihood fit.
The B, C, and D regions of the plane are not included in the likelihood function.
The analysis proceeds within the A region, where the ABCD background estimate is available.
The control region (CR) is defined to include bins in which $S / B < 0.2\%$ according to the predicted signal cross-section and the reweighted \EPOS MC\@.
The signal region is defined to include all bins where $S / B > 4\%$.
A likelihood validation region (LVR) is defined to include the bins where $0.2\% < S / B < 0.4\%$.
The LVR is used to test the extrapolation of the background model into bins that are not constrained in the fit, where the signal is negligible.
The remaining bins with $0.4\% < S / B < 4\%$ are assigned to a physics validation region (PVR).
The PVR and SR data are not looked at until the signal extraction strategy is finalised based on simulation\@.
The model is fitted to the CR and SR data.

The nominal background estimation is given by the ABCD prediction calculated from the data, denoted Data ABCD\@.
Differences to the three MC-based background models are included as correlated, one-sided NPs of the form $+(\text{\EPOS} - \text{Data ABCD})$, $+(\text{\PYTHIA{8}} - \text{Data ABCD})$, and $+(\text{\SHERPA{3}} - \text{Data ABCD})$.
The ABCD method can be applied to a MC model to get the corresponding MC ABCD prediction.
The non-closure of the ABCD method in \EPOS and \PYTHIA{8} constitutes further uncertainties, which are parameterized as the correlated one-sided NPs $+(\text{\EPOS} - \text{\EPOS ABCD})$ and $+(\text{\PYTHIA{8}} - \text{\PYTHIA{8} ABCD})$.
The non-closure in \EPOS is up to 20\% in the SR pre-fit.
The ABCD prediction made from the \SHERPA{3} prediction is not considered in the non-closure uncertainty since \SHERPA{3} disagrees with the data in the B, C, and D regions.
A background-only fit to the CR data produces a maximum non-closure of $~^{+2.3\%}_{-2.9\%}$ in the LVR\@.
This non-closure is included as an additional uncorrelated NP\@.
The fit prefers shifts of approximately $0.8\sigma$ towards the \SHERPA{3} and \EPOS ABCD non-closure NPs, with smaller shifts observed for the \EPOS, \PYTHIA{8} and \PYTHIA{8} ABCD non-closure NPs.

Signal theory uncertainties are quantified by varying the factorization and renormalization scales, \muF and \muR, by factors of two up and down.
The upper envelope of the scale uncertainty is from the $0.5 \times \muR$ variation, which increases the cross-section by a factor of about 24, while the downward part of the envelope is from the $0.5 \times \muF$ variation, which causes a reduction by a factor of about 20, occurring due to the reduction in the gluon density of the proton towards lower momentum transfers.
The uncertainty in the \CT[18nnlo] PDF set is included in quadrature with the \alphas = 0.117, 0.119 variations of the PDF fit and the alternative \NNPDF[3.0nnlo] PDF distribution.
The signal theory uncertainties only affect the normalization of the signal within the analysis selection, and are thus degenerate with the signal strength.
They are therefore not included in the likelihood fit, and the signal cross-section is rather constrained directly from the data in terms of $\mu$.
The uncertainty on the fitted signal strength is dominated by the non-closure of the ABCD method in \EPOS and \PYTHIA{8}, followed by the residual non-closure of the model with the data in the LVR.

The main results of the likelihood fit are given in the main text.
An alternative fit after correcting the ABCD plane for closure in \EPOS as opposed to the mean MC prediction yields a consistent signal strength of $\mu = 1.9 \pm 0.8$ with a significance of $3.0\sigma$, demonstrating that the conclusion is robust against the choice of closure-reweighting target.

\begin{figure}
\centering
\includegraphics[width=\textwidth]{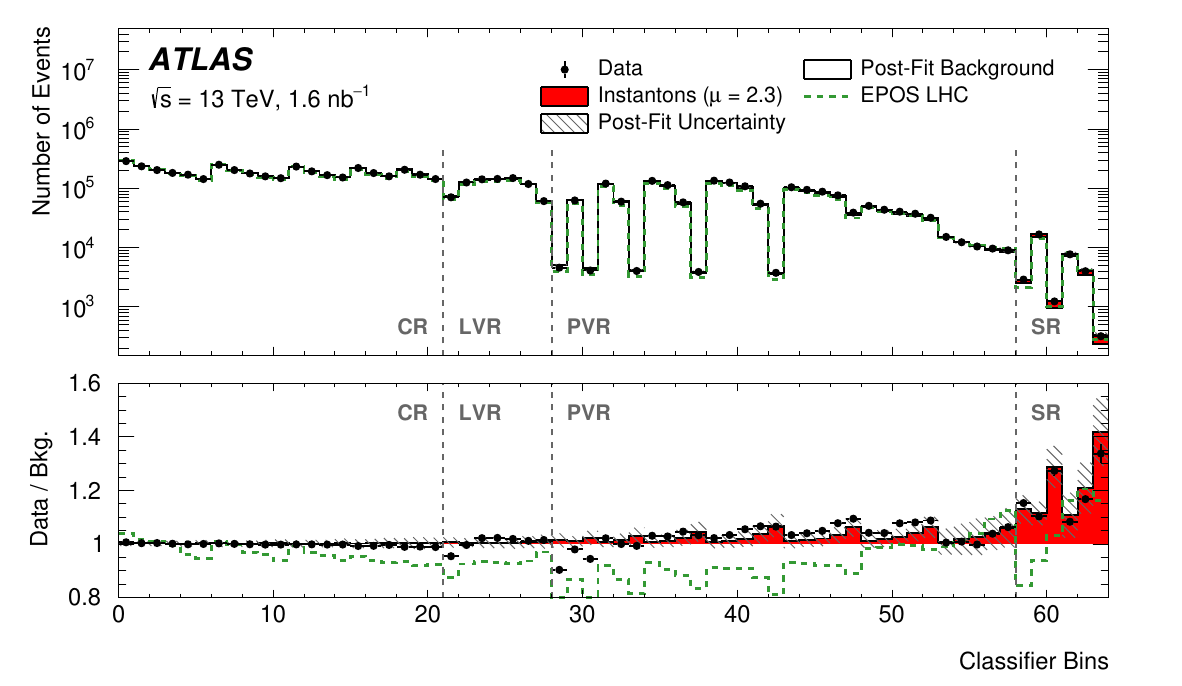}
\caption{Distribution of the unrolled classifier distribution post-fit.  The classifier bins are grouped into the control region (CR), signal region (SR), likelihood validation region (LVR), and the physics validation region (PVR), which was blinded together with the SR\@.  The post-fit background model shown as the white histogram comes from applying the post-fit pulls to the ABCD prediction.  The instanton signal scaled by the best-fit signal strength value of $\mu = 2.3 \pm 0.6$ is stacked on top.  The uncertainties in the post-fit pulls are added in quadrature and propagated through the ratio together with the uncertainty on the best-fit signal strength to form the hatched uncertainty band.  The data and the signal MC event yields are weighted by the ABCD non-closure correction factors.  The uncertainties on the data points are statistical only.  The reweighted \EPOS prediction is overlaid for comparison.}
\label{fig:classifier}
\end{figure}

In \Fig{\ref{fig:classifier}}, the bins of the A-region of the two-dimensional classifier plane are grouped into the likelihood regions and ordered by increasing classifier values.
The post-fit classifier distribution shows good agreement, within the post-fit uncertainties, between the fitted signal-plus-background model and the data across the entire classifier range.
The data in the PVR show a systematic offset above the model, but their distribution has the shape of the signal-plus-background model and remains within the post-fit uncertainties of the model throughout most of the bins.
This offset may be due to the contribution of lower-mass instanton signal, or from an underestimation of the background due to the large extrapolation between the CR to the SR\@.
It is verified that compatible results are obtained when the PVR is included in the likelihood function as part of the SR\@.

\FloatBarrier

\printbibliography
\clearpage
% ATLAS Collaboration author list
% Reference date of STDM-2021-01 is 2026-08-24
% Author list last updated on date 01-DEC-26
% Data extracted on 01-Dec-2026 for paper reference STDM-2021-01
% at 3:21pm
 
\begin{flushleft}
\hypersetup{urlcolor=black}
{\Large The ATLAS Collaboration}

\bigskip

\AtlasOrcid[0000-0002-6665-4934]{G.~Aad}$^\textrm{\scriptsize 102}$,
\AtlasOrcid[0000-0001-7616-1554]{E.~Aakvaag}$^\textrm{\scriptsize 17}$,
\AtlasOrcid[0000-0002-5888-2734]{B.~Abbott}$^\textrm{\scriptsize 121}$,
\AtlasOrcid[0000-0002-0287-5869]{S.~Abdelhameed}$^\textrm{\scriptsize 83b}$,
\AtlasOrcid[0000-0002-1002-1652]{K.~Abeling}$^\textrm{\scriptsize 54}$,
\AtlasOrcid[0000-0001-5763-2760]{N.J.~Abicht}$^\textrm{\scriptsize 48}$,
\AtlasOrcid[0000-0002-8496-9294]{S.H.~Abidi}$^\textrm{\scriptsize 30}$,
\AtlasOrcid[0009-0003-6578-220X]{M.~Aboelela}$^\textrm{\scriptsize 44}$,
\AtlasOrcid[0000-0002-9987-2292]{A.~Aboulhorma}$^\textrm{\scriptsize 36e}$,
\AtlasOrcid[0000-0001-5329-6640]{H.~Abramowicz}$^\textrm{\scriptsize 154}$
% ...
\bigskip
\\

% ...
$^{17}$Department for Physics and Technology, University of Bergen, Bergen; Norway.\\
% ...
$^{30}$Physics Department, Brookhaven National Laboratory, Upton NY; United States of America.\\
% ...
$^{36}$$^{(a)}$Facult\'e des Sciences Ain Chock, Universit\'e Hassan II de Casablanca;$^{(b)}$Facult\'{e} des Sciences, Universit\'{e} Ibn-Tofail, K\'{e}nitra;$^{(c)}$Facult\'e des Sciences Semlalia, Universit\'e Cadi Ayyad, LPHEA-Marrakech;$^{(d)}$LPMR, Facult\'e des Sciences, Universit\'e Mohamed Premier, Oujda;$^{(e)}$Facult\'e des sciences, Universit\'e Mohammed V, Rabat;$^{(f)}$Institute of Applied Physics, Mohammed VI Polytechnic University, Ben Guerir; Morocco.\\
% ...
$^{44}$Physics Department, Southern Methodist University, Dallas TX; United States of America.\\
% ...
$^{48}$Fakult\"{a}t Physik, Technische Universit{\"a}t Dortmund, Dortmund; Germany.\\
% ...
$^{54}$II. Physikalisches Institut, Georg-August-Universit\"{a}t G\"ottingen, G\"ottingen; Germany.\\
% ...
$^{83}$$^{(a)}$Khalifa University of Science and Technology, Abu Dhabi;$^{(b)}$New York University Abu Dhabi, Abu Dhabi;$^{(c)}$United Arab Emirates University, Al Ain;$^{(d)}$University of Sharjah, Sharjah; United Arab Emirates.\\
% ...
$^{102}$CPPM, Aix-Marseille Universit\'e, CNRS/IN2P3, Marseille; France.\\
% ...
$^{121}$Homer L. Dodge Department of Physics and Astronomy, University of Oklahoma, Norman OK; United States of America.\\
% ...
$^{154}$Raymond and Beverly Sackler School of Physics and Astronomy, Tel Aviv University, Tel Aviv; Israel.\\
% ...
$^{a}$ Also at Affiliated with an institute formerly covered by a cooperation agreement with CERN.\\
$^{b}$ Also at An-Najah National University, Nablus; Palestine.\\
$^{c}$ Also at Borough of Manhattan Community College, City University of New York, New York NY; United States of America.\\
$^{d}$ Also at Center for Interdisciplinary Research and Innovation (CIRI-AUTH), Thessaloniki; Greece.\\
$^{e}$ Also at Centre of Physics of the Universities of Minho and Porto (CF-UM-UP); Portugal.\\
$^{f}$ Also at CERN, Geneva; Switzerland.\\
% ...

\end{flushleft}

% Created with Glance <Atlas.Glance@cern.ch>

%

%
%

%

%
%
%
%

\end{document}